\documentclass[fleqn,twoside]{article}
\usepackage{espcrc2}
\usepackage{graphicx}

\readRCS
$Id: espcrc2.tex,v 1.2 2004/02/24 11:22:11 spepping Exp $
\newcommand{\ind}[2]{^{#1}_{\mbox{\scriptsize #2}}}
\newcommand{\al}[2]{\alpha\ind{#1}{#2}}
\newcommand{\MTau}{M_{\tau}}
\newcommand{\mpi}{m_{\pi}}
\newcommand{\DQCD}{\Delta_{\mbox{\tiny QCD}}}
\newcommand{\Nc}{N_{\mbox{\scriptsize c}}}
\newcommand{\nf}{n_{\mbox{\scriptsize f}}}
\newcommand{\Vud}{V_{\mbox{\scriptsize ud}}}
\newcommand{\Sew}{S_{\mbox{\tiny EW}}}
\newcommand{\dpew}{\delta'_{\mbox{\tiny EW}}}
\newcommand{\RTau}[1]{R_{\tau, \mbox{\tiny #1}}}

\title{On the low--energy behavior of the Adler function}

\author{A.V.~Nesterenko\address[JINR]{BLTPh, Joint Institute for Nuclear
Research, Joliot Curie 6, Dubna, 141980, Russian Federation}\thanks{E-mail:
nesterav@theor.jinr.ru}}

\begin{document}

\begin{abstract}
The infrared behavior of the Adler function is examined by making use of a
recently derived integral representation for the latter. The obtained
result for the Adler function agrees with its experimental prediction in
the entire energy range. The inclusive $\tau$~lepton decay is studied in
the framework of the developed approach.
\end{abstract}

\maketitle
\setcounter{footnote}{0}

\section{INTRODUCTION}

The Adler function~\cite{Adler} plays a key role in particle physics.
Specifically, theoretical description of some strong interaction processes
(e.g., electron--positron annihilation into hadrons~\cite{Feynman} and
inclusive $\tau$~lepton decay~\cite{BNP,Davier}) is inherently based on
this function. Besides, Adler function is essential for confronting the
precise experimental measurements of some electroweak observables (e.g.,
muon anomalous magnetic moment~\cite{muon} and shift of the
electromagnetic fine structure constant~\cite{Jegerlehner08}) with their
theoretical predictions. In turn, the latter represents a decisive test of
the Standard Model and imposes strict restrictions on possible ``new
physics'' beyond it.

Furthermore, Adler function plays a crucial role for the congruous
analysis of spacelike and timelike experimental data. Indeed, since
perturbation theory and renormalization group method are not applicable
directly to the study of observables depending on the timelike kinematic
variable, for the self--consistent description of the latter one has to
relate the timelike experimental data with the spacelike perturbative
results. Here, the required link between the experimentally measurable
$R$--ratio of electron--positron annihilation into hadrons and
theoretically computable Adler function~$D(Q^2)$ is represented by the
dispersion relation~\cite{Adler}
\begin{equation}
\label{AdlerDisp}
D(Q^2) = Q^2 \int_{4\mpi^2}^{\infty}
\frac{R(s)}{(s + Q^2)^2} \, d s,
\end{equation}
where $\mpi\simeq 135\,$MeV~\cite{PDG08} stands for the mass of the
lightest hadron state. The dispersion relation~(\ref{AdlerDisp}) is also
commonly employed for extracting the Adler function from the relevant
experimental data. For this purpose, in the integrand~(\ref{AdlerDisp})
$R(s)$ is usually parameterized by its experimental measurements at low
and intermediate energies and by its theoretical prediction at high
energies.

The ultraviolet behavior of the Adler function can be approximated by the
power series in the strong running coupling within the perturbation theory
(see paper~\cite{AdlerPert} and references therein)
\begin{equation}
\label{AdlerPert}
D^{(\ell)}_{\mbox{\scriptsize pert}}(Q^2) = 1 +
\sum\nolimits_{j=1}^{\ell} d_j\bigl[\al{(\ell)}{s}(Q^2)\bigr]^j.
\end{equation}
The overall factor $\Nc\sum_{f}Q_{f}^{2}$ is omitted throughout, where
$\Nc=3$ is the number of colors and $Q_{f}$ denotes the charge of the
quark of the $f$-th flavor. In Eq.~(\ref{AdlerPert}) $\al{(\ell)}{s}(Q^2)$
is the $\ell$--loop perturbative QCD invariant charge, $\al{(1)}{s}(Q^2) =
4\pi/(\beta_{0}\ln z)$, $z=Q^2/\Lambda^2$, $\beta_{0} = 11 - 2\nf/3$,
$\nf$ is the number of active quarks, and $d_{1}=1/\pi$.

However, the perturbative expansion~(\ref{AdlerPert}) is invalid at low
energies and it is inconsistent with the dispersion relation for the Adler
function~(\ref{AdlerDisp}) due to unphysical singularities of the strong
running coupling~$\al{}{s}(Q^2)$ in the infrared domain. The latter also
causes certain difficulties in processing the low--energy experimental
data.

\section{NOVEL INTEGRAL REPRESENTATION FOR THE ADLER FUNCTION}

In general, there is a variety of the nonperturbative approaches to handle
the strong interaction processes at low energies. In this work we will
focus on the approach which engages dispersion relations. Indeed, the
latter provide an important source of the nonperturbative information
about the hadron dynamics in the infrared domain, which should certainly
be taken into account when one is trying to go beyond the scope of
perturbation theory.

In particular, dispersion relation~(\ref{AdlerDisp}) imposes stringent
physical nonperturbative constraints on the Adler function. Specifically,
since $R(s)$, being the ratio of two cross--sections, assumes finite
values and tends to a constant in the ultraviolet asymptotic $s\to\infty$,
then the Adler function~$D(Q^2)$ vanishes\footnote{This constraint holds
for $\mpi \neq 0$ only.} in the infrared limit~$Q^2=0$. In addition,
dispersion relation~(\ref{AdlerDisp}) implies that the Adler function
possesses the only cut~$Q^2\le -4\mpi^2$ along the negative semi--axis of
real~$Q^2$.

These nonperturbative constraints on the Adler function have been merged
with its perturbative approximation in Refs.~\cite{AdlerIR,InPrep} (see
also discussion of this issue in Ref.~\cite{Prosperi}). Eventually, this
results in the following integral representations for the Adler function
and $R$--ratio:
\begin{eqnarray}
\label{AdlerInt}
D(Q^2)\!\!\!\! &=&\!\!\!\! \frac{Q^2}{Q^2+4\mpi^2}
\Biggl[1 + \!\!\!\int\limits_{4\mpi^2}^{\infty}\!\!\!
\rho(\sigma) \frac{\sigma - 4\mpi^2}{\sigma+Q^2}
\frac{d \sigma}{\sigma}\Biggr]\!, \\
\label{RInt}
R(s) &=& \theta(s-4\mpi^2) \Biggl[1 + \int\limits_{s}^{\infty}
\rho(\sigma)\,\frac{d \sigma}{\sigma}\Biggr],
\end{eqnarray}
where $\theta(x)$ is the unit step function, $\theta(x)=1$ if $x \ge 0$
and $\theta(x)=0$ otherwise. The developed approach~\cite{AdlerIR}
eliminates such intrinsic difficulties of perturbation theory as the
infrared unphysical singularities of outcoming results. Besides,
additional parameters are not introduced into the theory. Furthermore,
Eq.~(\ref{RInt}) by construction accounts for the effects due to the
analytic continuation of spacelike theoretical results into timelike
domain, such as the resummation of the so-called $\pi^2$--terms. It is
worth noting also that the mass of the lightest hadron state affects both
the parton model prediction and the strong correction of the quantities in
hand~(\ref{AdlerInt}),~(\ref{RInt}).

In the limit of the massless pion $\mpi=0$ expressions~(\ref{AdlerInt})
and (\ref{RInt}) become identical to those of the Analytic Perturbation
Theory (APT), see papers~\cite{APT1,APT2} and references therein. However,
it is crucial to keep the pion mass nonvanishing, since it can be safely
neglected only when one handles the strong interaction processes at high
energies. It is worth mentioning that there is a number of other similar
approaches\footnote{The Adler function was studied in the framework of APT
supplemented with the relativistic quark mass threshold resummation in
Ref.~\cite{MSS}.} which also combine perturbative results with relevant
dispersion relations, see, e.g.,
Refs.~\cite{BMS,Arbuzov,Maxwell,Fischer,Cvetic}.

\begin{figure}[t]
\centerline{\includegraphics[width=72.5mm,clip]{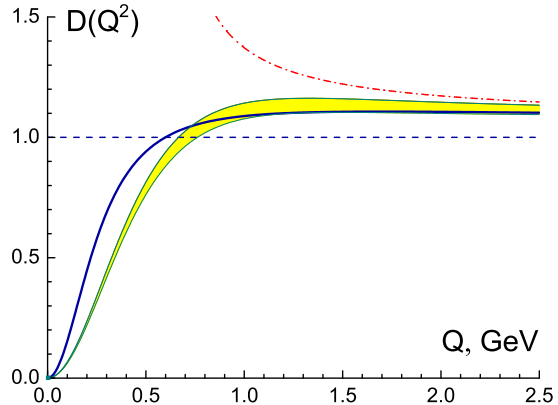}}
\vskip-2.5mm
\caption{Adler function~(\ref{AdlerInt}) corresponding to the spectral
density~(\ref{Rho}) (solid curve) ($\Lambda=441\,$MeV, $\nf=3$).
Perturbative approximation~(\ref{AdlerPert}) of the Adler function and its
experimental prediction are denoted by the dot-dashed curve and shaded
band, respectively.}
\label{Plot:AdlerIR}
\end{figure}

The spectral density~$\rho(\sigma)$, which appears in
Eqs.~(\ref{AdlerInt}) and~(\ref{RInt}), can be determined either as the
discontinuity of the explicit ``exact'' theoretical expression for the
Adler function~$D_{\mbox{\scriptsize{exact}}}(Q^2)$ across the physical
cut or as the numerical derivative of the experimental data on
R--ratio~\cite{AdlerIR}:
\begin{equation}
\label{RhoExact}
\rho(\sigma) = \frac{1}{\pi}\,
D_{\mbox{\scriptsize{exact}}}(-\sigma + i 0_{+}) =
-\frac{d\,R_{\mbox{\scriptsize{exp}}}(\sigma)}{d\,\ln\sigma}.
\end{equation}
However, there is still no explicit ``exact'' expression for the Adler
function, and, therefore, there is no unique way to compute the
corresponding spectral density~(\ref{RhoExact}) by making use of its
approximate perturbative expression~(\ref{AdlerPert}). In what follows we
will employ the spectral function obtained in Ref.~\cite{PRD}\footnote{It
is interesting to note that the QCD effective coupling obtained in
Ref.~\cite{PRD} has been independently rediscovered in
Ref.~\cite{Schrempp} proceeding from entirely different reasoning.}, which
has the following form at the one-loop level~\cite{PRD,Review}:
\begin{equation}
\label{Rho}
\rho^{(1)}(\sigma) = \left(1+\frac{\Lambda^2}{\sigma}\right)
\frac{1}{\ln^2(\sigma/\Lambda^2) + \pi^2}.
\end{equation}
The Adler function~(\ref{AdlerInt}), corresponding to the spectral
function~(\ref{Rho}), is presented in Fig.~\ref{Plot:AdlerIR} by solid
curve. The dot-dashed curve stands for the one-loop perturbative
approximation~(\ref{AdlerPert}) of the Adler function, whereas its
experimental prediction, computed in the way described above, is denoted
by the shaded band. As one may infer from Fig.~\ref{Plot:AdlerIR}, the
obtained result for the Adler function is in a reasonable agreement with
its experimental prediction in the entire energy range.

\section{INCLUSIVE {\Large $\tau$}~LEPTON DECAY}

It is also of a particular interest to study the inclusive $\tau$~lepton
decay within the approach in hand, since this process probes the infrared
hadron dynamics at energies below the mass of the $\tau$~lepton, and the
relevant experimental data are fairly precise. The measurable quantity
here is the inclusive semileptonic branching ratio
\begin{equation}
\label{RTauDef}
R_{\tau} =
\frac{\Gamma(\tau^{-} \to \mbox{hadrons}^{-}\, \nu_{\tau})}
{\Gamma(\tau^{-} \to e^{-}\, \bar\nu_{e}\, \nu_{\tau})},
\end{equation}
which can be split into three parts, namely, $R_{\tau}= \RTau{V} +
\RTau{A} + \RTau{S}$. The terms $\RTau{V}$ and $\RTau{A}$ account for the
contributions to Eq.~(\ref{RTauDef}) of the decay modes with the light
quarks only, and they correspond to the vector (V) and axial--vector~(A)
quark currents, respectively. The last term $\RTau{S}$ accounts for the
contribution to Eq.~(\ref{RTauDef}) of the decay modes with the $s$~quark.

Let us proceed with the nonstrange part of the ratio~(\ref{RTauDef})
associated with the vector quark currents
\begin{equation}
\label{RTauV}
\RTau{V} = \frac{\Nc}{2}\,|\Vud|^2 \Sew
\left(\DQCD + \dpew \right),
\end{equation}
see papers~\cite{BNP,Davier} and references therein for detailed
discussion of this issue. The experimental measurement~\cite{TauExp} of
the ratio~(\ref{RTauV}) yields $\RTau{V}= 1.764 \pm 0.016$. In
Eq.~(\ref{RTauV}) $|\Vud|=0.97418 \pm 0.00027$ denotes the
Cabibbo--Kobayashi--Maskawa matrix element~\cite{PDG08}, $\Sew = 1.0194
\pm 0.0050$ and $\delta'_{\mbox{\tiny EW}}=0.0010$ are the electroweak
corrections~\cite{BNP,EWF}, and $\DQCD$ can be expressed in terms of a
weighted integral of the aforementioned $R(s)$--ratio:
\begin{equation}
\label{DQCD}
\DQCD = 2\!\!\int\limits_{0}^{\MTau^2}\!\!
\left(1 - \frac{s}{\MTau^2}\right)^{\!\!2}\!\!
\left(1 + 2\, \frac{s}{\MTau^2}\right)\! R(s)
\frac{d s}{\MTau^2},
\end{equation}
where $\MTau\simeq 1.777\,$GeV~\cite{PDG08} is the $\tau$~lepton mass.

In the framework of perturbative approach one usually reduces
Eq.~(\ref{DQCD}) to the contour integral in the complex $s$--plane along
the circle of the radius of the squared mass of the $\tau$~lepton. At the
one-loop level this eventually leads to~\cite{BNP}
\begin{equation}
\DQCD = 1 + d_1\,\al{(1)}{s}(\MTau^2),
\end{equation}
that, in turn, results in $\Lambda = (678 \pm 55)\,$MeV for $\nf=2$ active
quarks.

At the same time, for the evaluation of $\DQCD$ in the framework of the
approach in hand, the integration in Eq.~(\ref{DQCD}) can be performed in
a straightforward way. Ultimately this leads to the following result at
the one-loop level~\cite{InPrep}:
\begin{eqnarray}
\label{DQCDMAPT}
\DQCD\!\!\!\!&=&\!\!\!\! 1 - \delta_{\Gamma} +
d_1 \al{(1)}{\tiny TL}(\MTau^2) -
d_1 \delta_{\Gamma} \al{(1)}{\tiny TL}(m_{\Gamma}^{2}) \nonumber \\
&+&\!\!\!\!
d_1\,\frac{4\pi}{\beta_{0}}\int_{\chi}^{1}\! f(\xi)\,
\rho^{(1)}(\xi \MTau^2)\, d \xi,
\end{eqnarray}
where $f(\xi)=\xi^3-2\xi^2+2$, $\chi=m_{\Gamma}^2/\MTau^2$,
$\delta_{\Gamma}=\chi\,f(\chi)$, and
\begin{equation}
\al{(1)}{\tiny TL}(s) = \frac{4 \pi}{\beta_0}\,\theta(s-m_{\Gamma}^{2})\!
\int_{s}^{\infty} \rho^{(1)}(\sigma)\,\frac{d \sigma}{\sigma}
\end{equation}
is the one-loop timelike effective coupling~\cite{AdlerIR}. Here
$m_{\Gamma}$ stands for the total mass of the lightest allowed hadronic
decay mode of the $\tau$~lepton, e.g., for the vector channel $m_{\Gamma}
= m_{\pi^0} + m_{\pi^{-}}$. In this case $\delta_{\Gamma} \simeq 0.048$
considerably exceeds the electroweak correction~$\delta'_{\mbox{\tiny
EW}}$. Eventually, Eq.~(\ref{DQCDMAPT}) results in $\Lambda=(941 \pm
86)\,$MeV for $\nf=2$ active quarks, that is somewhat larger than the
one-loop perturbative estimation quoted above.

The effects due to the nonvanishing hadronic mass~$m_{\Gamma}$ play a
substantial role herein. In particular, in the massless limit
$m_{\Gamma}=0$ Eq.~(\ref{DQCDMAPT}) leads to $\Lambda=(493 \pm 56)\,$MeV
for $\nf=2$ active quarks.

\section{SUMMARY}

The infrared behavior of the Adler function is studied by making use of
recently derived integral representation for the latter. The developed
approach possesses a number of appealing features. Namely, it eliminates
unphysical perturbative singularities, properly accounts for the effects
due to the analytic continuation of spacelike theoretical results into
timelike domain, and embodies the effects due to the mass of the lightest
hadron state. Besides, additional adjustable parameters are not introduced
into the theory. Furthermore, the developed approach provides a reasonable
description of the Adler function in the entire energy range. It is also
shown that the effects due to the nonvanishing mass of the lightest hadron
state play a substantial role in processing the experimental data on the
inclusive $\tau$~lepton decay.

\vskip2mm
\noindent
\textbf{Acknowledgements}
\vskip1mm

This work was partially performed during the visit of the author to the
University of Milano. Author is thankful to Professor Giovanni Prosperi
for his kind hospitality and fruitful discussions. Author is grateful to
N.~Brambilla, S.~Forte, and A.~Vairo for the interest to this study.
Partial financial support of grants RFBR-08-01-00686, BRFBR-JINR-F08D-001,
and NS-1027.2008.2 is acknowledged.

\end{document}